\documentclass[twocolumn,apj]{aastex62}
\usepackage{lineno}

\begin{document}

\title{Spectroscopic orbits of subsystems in  multiple
  stars. IV. Double-lined pairs.}

\author{Andrei Tokovinin}
\affiliation{Cerro Tololo Inter-American Observatory, Casilla 603, La Serena, Chile}
\email{atokovinin@ctio.noao.edu}

\begin{abstract}
Spectroscopic orbits  are computed for  inner pairs in  9 hierarchical
multiple systems (HIP 19639, 60845, 75663, 76816, 78163, 78416, 80448,
84789,  and  HD~105080) and  for  one  simple  binary HIP~61840.   All
subsystems  are double-lined, and their  periods range  from 2.27  to 30.4
days. Five  spectroscopic binaries are  twins with equal  masses. Each
hierarchical system is  discussed individually, providing estimates of
outer periods,  masses, orbital inclination, and  axial rotation.  For
systems with  three resolved visual components (HIP  60845 and 80448),
the outer  and inner visual  orbits are determined,  complementing the
description   of  their  architecture.    The  radial   velocities  of
HIP~75663A, 76816B,  and 78163B are  found to be  variable with
long periods, implying that these hierarchies are 2+2 quadruples.  The
period-eccentricity   relation   for   spectroscopic   subsystems   is
discussed.
   \keywords{binaries:spectroscopic, binaries:visual}
\end{abstract}


\section{Introduction}
\label{sec:intro}

This  paper continues  the  series on  spectroscopic  orbits of  stars
belonging to hierarchical  systems \citep{paper1,paper2,paper3}. It is
motivated by  the need  to improve statistics  of orbital  elements in
stellar  hierarchies. Statistics will  inform us  on the  processes of
their formation  and dynamical evolution, as outlined  in the previous
papers  of  this  series.    This  work  augments  the  collection  of
observational data  on stellar  hierarchies assembled in  the multiple
star catalog \citep[MSC;][]{MSC}.

\begin{deluxetable*}{c c rr   l cc rr r c }
\tabletypesize{\scriptsize}     
\tablecaption{Basic parameters of observed multiple systems
\label{tab:objects} }  
\tablewidth{0pt}                                   
\tablehead{                                                                     
\colhead{WDS} & 
\colhead{Comp.} &
\colhead{HIP} & 
\colhead{HD} & 
\colhead{Spectral} & 
\colhead{$V$} & 
\colhead{$V-K$} & 
\colhead{$\mu^*_\alpha$} & 
\colhead{$\mu_\delta$} & 
\colhead{RV} & 
\colhead{$\overline{\omega}$\tablenotemark{a}} \\
\colhead{(J2000)} & 
 & &   &  
\colhead{Type} & 
\colhead{(mag)} &
\colhead{(mag)} &
\multicolumn{2}{c}{ (mas yr$^{-1}$)} &
\colhead{(km s$^{-1}$)} &
\colhead{(mas)} 
}
\startdata
04125$-$3609 &A & 19639 & 26758 & F3V     & 7.12 & 1.12 & 61   &   24  & 35.40    & 7.94 \\
             &B & 19646 & 26773 & F2IV    & 7.91 & 0.92 & 61   &   24  & 35.98    & 8.03 \\
12059$-$4951 &A & \ldots & 105080 & G3V   & 9.13 & 1.43 & 25   &$-$15  & 50.19    & 9.99 \\
             &B & \ldots & 105081 & G0V   & 9.18 & 1.42 & 31   &$-$20  & 50.09    & 7.20 \\
12283$-$6146 &A & 60845  & 108500 & G3V   & 6.82 & 1.64 & 71   &$-$160 & 40.02    & 19.93 \\
             &D & \ldots & \ldots & \ldots& 13.70 & 4.54& 73   &$-$169 & \ldots    & 19.91 \\
12404$-$4924 &A & 61840  & 110143 & G0V & 7.60 & 2.00 & $-$28 & $-$112 & 6.76     & 18.59 \\  
15275$-$1058 &A & 75663  & 137631 & G0    & 8.14  & 1.35&$-$65 &$-$35  & $-$56.0 v & 9.29 \\
             &B & \ldots & \ldots & G0    & 9.21  & 1.50&$-$61 &$-$35  & $-$56.82 & 7.69 \\
15410$-$1449 &A & 76816  & 139864 & F8V   & 9.47  & 1.62&$-$26 &$-$1   & $-$38.94 & 3.23 \\
             &B & \ldots & \ldots & \ldots& 9.74 & 2.51&$-$25  &$-$2   & $-$50.9 v & 3.15 \\  
15577$-$3915 &A & 78163  & 142728 & G3V   & 9.04 & 1.54& 17    & 6     & 9.41     & 10.42 \\
             &B & \ldots & \ldots & \ldots&10.30 & \ldots& 31  & 4     & 6.78 v     & 13.57 \\
16005$-$3605 &A & 78416  & 143215 & G1V   &8.65 & 1.32 &$-$26  &$-$41  & 1.60     & 9.33 \\
             &B & \ldots & \ldots & \ldots&9.32 & 1.31 &$-$28  &$-$41  & 1.43     & 9.31 \\
16253$-$4909 &AB& 80448  & 147633 & G2V   & 7.5? &2.3? &$-$95  &$-$94  & $-$2.08  & 19.66 \\
17199$-$1121 &A & 84789  & 156769 & F2    & 9.11 & 1.37&   6   & 13    & 5.62    & 5.33 \\
             &B & \ldots & \ldots & \ldots& 9.89 & 1.37&  5    & 12    & 5.97    & 5.34    
\enddata
\tablenotetext{a}{Proper motions and parallaxes are 
  from the {\it Gaia} DR2 \citep{Gaia,Gaia1}.}
\end{deluxetable*}

The  systems studied  here are  presented  in Table~\ref{tab:objects}.
Only one of them (HIP 61840) is a simple binary belonging to the 67 pc
sample of solar-type stars; others  contain from three to five components and
are also relatively  close to the Sun. Their  principal components are
main sequence stars with spectral types  from F2V to G3V.  The data in
Table~\ref{tab:objects} are collected from  Simbad and {\it Gaia} DR2 \citep{Gaia},
the  radial velocities  (RVs) are  determined here  (variable  RVs are
marked by 'v').

The structure of this paper is similar to the previous ones.  The data
and methods  are briefly recalled in  Section~\ref{sec:obs}, where the
new  orbital elements  are also  given. Then  in Section~\ref{sec:obj}
each system is  discussed individually. The paper closes  with a short
summary in Section~\ref{sec:sum}.

\section{Observations and data analysis}
\label{sec:obs}

\subsection{Spectroscopic observations}

The spectra used here were taken with the 1.5 m telescope sited at the
Cerro Tololo  Inter-American Observatory (CTIO) in  Chile and operated
by             the             SMARTS            Consortium.\footnote{
  \url{http://www.astro.yale.edu/smarts/}}  The   observing  time  was
allocated  through  NOAO.   Observations  were made  with  the  CHIRON
optical echelle spectrograph \citep{CHIRON} by the telescope operators
in service mode.  In two runs,  2017 August and 2018 March, the author
also observed in classical mode.  All spectra are taken in the slicer
mode with a resolution of $R=80,000$ and a signal to noise ratio of at
least 20.  Thorium-Argon calibrations were recorded for each target.

Radial  velocities are  determined  from the  cross-correlation
function (CCF) of echelle orders with the binary mask based on the solar
spectrum,  as detailed  in  \citep{paper1}. The  RVs  derived by  this
method should be  on the absolute scale if  the wavelength calibration
is  accurate.  The  CHIRON  RVs were  checked  against standards  from
\citep{Udry1998}, and a small  offset of $+0.15$ km~s$^{-1}$ was found
in  \cite{paper3}.   

The CCF contains two dips  in the case of double-lined systems studied
here.  The  dip width  is related to  the projected rotation  speed $V
\sin i$, while its area depends on the spectral type, metallicity, and
relative  flux.  Table~\ref{tab:dip} lists  average parameters  of the
Gaussian  curves fitted  to  the CCF  dips.   It gives  the number  of
averaged  measurements  $N$ (blended  CCFs  were  not  used), the  dip
amplitude  $a$,  its  dispersion  $\sigma$,  the  product  $a  \sigma$
proportional to  the dip  area (hence to  the relative flux),  and the
projected rotation velocity $V \sin i$, estimated from $\sigma$ by the
approximate  formula   given  in  \citep{paper1}.    The  last  column
indicates  the presence  or absence  of the  lithium 6708\AA  ~line in
individual components.

\begin{deluxetable*}{l l c cccc c}    
\tabletypesize{\scriptsize}     
\tablecaption{CCF parameters
\label{tab:dip}          }
\tablewidth{0pt}                                   
\tablehead{                                                                     
\colhead{HIP/HD} & 
\colhead{Comp.} & 
\colhead{$N$} & 
\colhead{$a$} & 
\colhead{$\sigma$} & 
\colhead{$a \sigma$} & 
\colhead{$V \sin i$ } & 
\colhead{Li}
\\
 &  &  & &
\colhead{(km~s$^{-1}$)} &
\colhead{(km~s$^{-1}$)} &
\colhead{(km~s$^{-1}$)} &
\colhead{  6708\AA}
}
\startdata
HIP 19639  & Aa &  7 & 0.061   &  12.53  &    0.765 &   21.7 &N\\
HIP 19639  & Ab &  7 & 0.033   &   5.90  &    0.193 &    8.7 &N\\
HIP 19646  & B  &  2 & 0.042    &  18.87  &    0.787 &   33: &N\\ 
HD 105080  & A  &  3 & 0.389   &   3.68  &    1.429 &    2.5 &N\\ 
HD 105081  & Ba & 11 & 0.179   &   3.82  &    0.684 &    3.1 &N\\
HD 105081  & Bb & 11 & 0.167   &   3.78  &    0.630 &    3.0 &N\\
HIP 60845  & Aa &  2 & 0.183   &   3.83  &    0.702 &    3.2 &N\\
HIP 60845  & Ab &  2 & 0.124   &   4.01  &    0.497 &    3.8 &N\\
HIP 60845  & BC &  2 & 0.430   &   3.57  &    1.535 &    2.0 &N\\ 
HIP 61840  & Aa &  9 & 0.189   &   4.51  &    0.853 &    5.3 &Y\\
HIP 61840  & Ab &  9 & 0.124   &   4.13  &    0.511 &    4.2 &Y\\
HIP 75663  & A  &  5 & 0.223   &   6.85  &    1.528 &   10.7 &Y\\
HIP 75663  & Ba & 12 & 0.161   &   4.88  &    0.789 &    6.3 &Y\\ 
HIP 75663  & Bb & 12 & 0.155   &   4.87  &    0.754 &    6.3 &Y\\
HIP 76816  & Aa &  6 & 0.110   &   8.14  &    0.899 &   13.3 &Y\\ 
HIP 76816  & Ab &  6 & 0.030   &   4.49  &    0.137 &    5.3 &Y\\
HIP 76816  & B  &  5 & 0.506   &   3.78  &    1.913 &    3.0 &Y\\
HIP 78163  & Aa &  9 & 0.197   &   4.08  &    0.803 &    4.0 &Y\\
HIP 78163  & Ab &  9 & 0.192   &   4.05  &    0.778 &    4.0 &Y\\
HIP 78163  & B  &  3 & 0.371   &   4.67  &    1.732 &    5.8 &N\\ 
HIP 78416  & Aa &  9 & 0.047   &  15.43  &    0.725 &   27: &Y\\
HIP 78416  & Ab &  9 & 0.041   &  13.18  &    0.536 &   23: &Y\\
HIP 78416  & B  &  4 & 0.058   &  22.68  &    1.324 &   40: &Y\\ 
HIP 80448  & Aa &  2 & 0.101   &  12.30  &    1.247 &   21.3 &Y\\ 
HIP 80448  & Ab &  2 & 0.023   &   9.78  &    0.221 &   16.5 &Y\\
HIP 80448  & B  &  3 & 0.171   &   8.62  &    1.469 &   14.3 &Y\\    
HIP 84789  & Aa &  6 & 0.043   &  12.41  &    0.536 &   21.5 &N\\
HIP 84789  & Ab &  6 & 0.048   &  10.88  &    0.522 &   18.6 &N\\
HIP 84789  & B  &  2 & 0.036   &  27.49  &    0.996 &   49: & N
\enddata 
\end{deluxetable*}

\subsection{Speckle interferometry}

Information  on   the  resolved  subsystems  is   retrieved  from  the
Washington Double Star Catalog \citep[WDS;][]{WDS}. It is complemented
by  recent  speckle   interferometry  at  the  Southern  Astrophysical
Research  (SOAR)  telescope.   The  latest  publication  \citep{SAM18}
contains references to previous papers.

\subsection{Orbit calculation}

As in Paper  3 \citep{paper3}, orbital elements and  their errors are
determined   by  the   least-squares  fits   with   weights  inversely
proportional   to   the   adopted   errors.    The   IDL   code   {\tt
  orbit} \citep{ORBIT}\footnote{Codebase: \url{http://www.ctio.noao.edu/~atokovin/orbit/} and 
\url{http://dx.doi.org/10.5281/zenodo.61119} }
is   used.   It   can   fit  spectroscopic,   visual,   or   combined
visual/spectroscopic  orbits. Formal  errors of  orbital  elements are
determined from these fits.   The elements of spectroscopic orbits are
given in Table~\ref{tab:sborb} in common notation.

\begin{deluxetable*}{l l cccc ccc c c}    
\tabletypesize{\scriptsize}     
\tablecaption{Spectroscopic orbits
\label{tab:sborb}          }
\tablewidth{0pt}                                   
\tablehead{                                                                     
\colhead{HIP/HD} & 
\colhead{System} & 
\colhead{$P$} & 
\colhead{$T$} & 
\colhead{$e$} & 
\colhead{$\omega_{\rm A}$ } & 
\colhead{$K_1$} & 
\colhead{$K_2$} & 
\colhead{$\gamma$} & 
\colhead{rms$_{1,2}$} &
\colhead{$M_{1,2} \sin^3 i$} 
\\
& & \colhead{(days)} &
\colhead{(+24,00000)} & &
\colhead{(degree)} & 
\colhead{(km~s$^{-1}$)} &
\colhead{(km~s$^{-1}$)} &
\colhead{(km~s$^{-1}$)} &
\colhead{(km~s$^{-1}$)} &
\colhead{ (${\cal M}_\odot$) } 
}
\startdata
HIP 19639 & Aa,Ab & 2.35254 & 58002.5732 & 0.0 & 0.0          & 39.474 & 49.194 & 35.400 & 0.36 & 0.094 \\
         && $\pm$0.00005 & $\pm$0.0019 & fixed & fixed & $\pm$0.160 & $\pm$0.298 & $\pm$0.100 & 0.87 & 0.076 \\
HD 105081 & Ba,Bb & 30.427 & 58214.038 & 0.418 & 82.1         & 46.856 & 47.493 & 50.099 & 0.07 & 1.00 \\
         && $\pm$0.006 & $\pm$0.023 & $\pm$0.001 & $\pm$0.3 & $\pm$0.117 & $\pm$0.118 & $\pm$0.041 & 0.05 & 0.98 \\
HIP 60845 & Aa,Ab & 6.3035 & 58195.6279 & 0.0 & 0.0           & 31.361 & 32.128 & 40.018 & 0.10 & 0.084 \\
         && $\pm$0.0001 & $\pm$0.0017 & fixed & fixed & $\pm$0.063 & $\pm$0.114 & $\pm$0.036 & 0.27 & 0.082\\
HIP 61840 & Aa,Ab& 9.6717 & 58194.383 & 0.007 & 357.0         & 56.665 & 60.744 & 6.745 & 0.05 & 0.84 \\
         && $\pm$0.0008 & $\pm$0.345 & $\pm$0.001 & $\pm$13.0 & $\pm$0.525 & $\pm$0.562 & $\pm$0.048 & 0.05 & 0.78\\
HIP 75663 & Ba,Bb &  22.8704 & 58204.4963 & 0.613 & 260.1     & 49.001 & 49.622 & $-$56.847 & 0.29 & 0.56 \\
         &&  $\pm$0.0047 & $\pm$0.020 & $\pm$0.001 & $\pm$0.3 & $\pm$0.126 & $\pm$0.134 & $\pm$0.045 & 0.27 & 0.56\\
HIP 76816 & Aa,Ab & 6.95176 & 58197.3528 & 0.0 & 0.0          & 45.790 & 62.448 & $-$39.145 & 0.29 & 0.53 \\
         && $\pm$0.00002 & $\pm$0.0063 & fixed & fixed & $\pm$0.306 & $\pm$0.570 & $\pm$0.186 & 1.13 & 0.39 \\
HIP 78163 & Aa,Ab &  21.8186 & 58202.091 & 0.577 & 301.0      & 45.429 & 45.671 & 9.411 & 0.04 & 0.47\\
         &&  $\pm$0.0015 & $\pm$0.014 & $\pm$0.002 & $\pm$0.3 & $\pm$0.161 & $\pm$0.186 & $\pm$0.045 & 0.04 & 0.46\\
HIP 78416 & Aa,Ab & 21.0802 & 58197.2344 & 0.708 & 99.1       & 64.636 & 71.737 & 1.636 &  0.16 & 1.03 \\
         &&  $\pm$0.0030 & $\pm$0.0072 & $\pm$0.003 & $\pm$0.3 & $\pm$0.372 & $\pm$0.509 & $\pm$0.112 & 0.61 & 0.93\\
HIP 80448 &Aa,Ab  & 2.2699 & 58195.5948 & 0.0 & 0.0           & 73.124 & 108.452 & $-$1.848 & 0.64 & 0.84 \\
         && $\pm$0.0002 & $\pm$0.0033 & fixed & fixed & $\pm$0.592 & $\pm$0.698 & $\pm$0.270 & 0.51 & 0.57\\
HIP 84789 &Aa,Ab & 2.2758 & 58196.7968 & 0.0 & 0.0            & 78.044 & 79.004 & 5.624 & 0.40 & 0.46 \\
         && $\pm$0.0001 & $\pm$0.0020 & fixed & fixed & $\pm$0.181 & $\pm$0.182 & $\pm$0.073 & 0.21 & 0.45
\enddata 
\end{deluxetable*}

\begin{deluxetable*}{l l cccc ccc}    
\tabletypesize{\scriptsize}     
\tablecaption{Visual orbits
\label{tab:vborb}          }
\tablewidth{0pt}                                   
\tablehead{                                                                     
\colhead{HIP/HD} & 
\colhead{System} & 
\colhead{$P$} & 
\colhead{$T$} & 
\colhead{$e$} & 
\colhead{$a$} & 
\colhead{$\Omega_{\rm A}$ } & 
\colhead{$\omega_{\rm A}$ } & 
\colhead{$i$ }  \\
& & \colhead{(years)} &
\colhead{(years)} & &
\colhead{(arcsec)} & 
\colhead{(degree)} & 
\colhead{(degree)} & 
\colhead{(degree)} 
}
\startdata
HD 105080 & Aa,Ab  & 91.6       & 1999.15   & 0.40      & 0.176             &  5.4      & 224.1    & 40.4  \\
HIP 60845 &   A,BC  & 690        & 1826.7    & 0.20      & 2.485             & 104.4    & 156.6      &  141.4  \\
HIP 60845 &   B,C   & 28.2      & 1990.2     & 0.166     & 0.221             & 162.8    & 77.3       & 156.1  \\
      &         & $\pm$0.2  & $\pm$0.3 & $\pm$0.009  & $\pm$0.003        & $\pm$6.5  & $\pm$5.9  & $\pm$2.5  \\
HIP 80448 &   A,B   & 1950      & 1926.58    & 0.63      & 4.644             & 75.5     & 267.9      & 152.1 \\
HIP 80448 &   Ba,Bb & 20.0      & 2018.08    & 0.42      & 0.176             & 6.1      & 150.9      & 109.1  \\
      &         & $\pm$0.3  & $\pm$0.24 & $\pm$0.02  & $\pm$0.003        & $\pm$1.2  & $\pm$4.9  & $\pm$0.8  
\enddata 
\end{deluxetable*}

For two multiple systems, the resolved measurements of inner and outer
pairs are represented by  visual orbits. Simultaneous fitting of inner
and outer orbits is done  using the code {\tt orbit3.pro} described by
\citet{TL2017}; the code is available in \citep{ORBIT3}.\footnote{Codebase: \url{http://dx.doi.org/10.5281/zenodo.321854}}
It accounts for the wobble in  the trajectory of the outer pair caused
by the subsystem.  The wobble amplitude is $f$  times smaller than the
inner   semimajor  axis,  where   the  wobble   factor  $f   =  q_{\rm
  in}/(1+q_{\rm in})$  depends on the  inner mass ratio  $q_{\rm in}$.
The elements  of visual orbits are given  in Table~\ref{tab:vborb}. As
outer orbits are  poorly constrained, I do not  list their errors. The
outer  orbits  serve primarily  to  model  the  observed part  of  the
trajectory for the determination  of $f$.  In the figures illustrating
these orbits,  the observed trajectories  are plotted relative  to the
primary component of each system, on the same scale.

Individual RVs  of spectroscopic binaries  and their residuals  to the
orbits are presented  in Table~\ref{tab:rv}. The HIP or  HD number and
the system  identifier (components joined  by comma) in the  first two
columns define the binary. Then follow  the Julian date, the RV of the
primary  component $V_1$,  its adopted  error $\sigma_1$  (blended CCF
dips  are assigned large  errors), and  its residual  (O$-$C)$_1$. The
last  three columns  give  velocities, errors,  and  residuals of  the
secondary  component. Table~\ref{tab:rvconst}  contains  RVs of  other
components,     both      constant     and     variable.      Finally,
Table~\ref{tab:speckle}  lists  position  measurements  used  for  the
calculation of visual orbits. It contains the HIP or HD number, system
identification,   date  of   observation,  position   angle  $\theta$,
separation  $\rho$,  adopted  error  $\sigma_\rho$ (errors  in  radial  and
tangential directions  are considered to be equal),  and the residuals
to the  orbits in $\theta$ and  $\rho$. The last  column indicates the
measurement technique.   Measurements of the outer systems  are of two
kinds: when the inner pair  is unresolved (e.g.  HIP 60845 A,BC), they
refer  to   the  photo-center  of  the  inner   pair,  while  resolved
measurements  refer  to the  individual  components  (e.g.  HIP  60845
A,B). The orbit-fitting code  accounts for reduced wobble amplitude of
unresolved (photo-center) measurements compared to resolved ones.


\begin{deluxetable*}{r l c rrr rrr}    
\tabletypesize{\scriptsize}     
\tablecaption{Radial velocities and residuals (fragment)
\label{tab:rv}          }
\tablewidth{0pt}                                   
\tablehead{                                                                     
\colhead{HIP/HD} & 
\colhead{System} & 
\colhead{Date} & 
\colhead{$V_1$} & 
\colhead{$\sigma_1$} & 
\colhead{(O$-$C)$_1$ } &
\colhead{$V_2$} & 
\colhead{$\sigma_2$} & 
\colhead{(O$-$C)$_2$ } \\
 & & 
\colhead{(JD $+$24,00000)} &
\multicolumn{3}{c}{(km s$^{-1}$)}  &
\multicolumn{3}{c}{(km s$^{-1}$)}  
}
\startdata
 19639 & Aa,Ab &  57985.8910 &  68.74 &   0.30 &   0.16  &    $-$6.37 &   0.60 &  $-$0.42   \\
 19639 & Aa,Ab &  57986.8980 &  15.11 &   0.30 &   0.18  &    59.76 &   0.60 &  $-$1.15   \\
 19639 & Aa,Ab &  58052.6260 &  31.22 &  10.00 &   2.23  & \ldots   &\ldots  & \ldots   \\ 
 19639 & Aa,Ab &  58053.6080 &  22.97 &   0.50 &   1.26  &    54.35 &   1.00 &   1.89   
\enddata 
\end{deluxetable*}

\begin{deluxetable}{r l r r }    
\tabletypesize{\scriptsize}     
\tablecaption{Radial velocities of other components
\label{tab:rvconst}          }
\tablewidth{0pt}                                   
\tablehead{                                                                     
\colhead{HIP/HD} & 
\colhead{Comp.} & 
\colhead{Date} & 
\colhead{RV}   \\ 
 & & 
\colhead{(JD $+$24,00000)} &
\colhead {(km s$^{-1}$)}  
}
\startdata
19646  &  B   & 57985.8932&  36.005 \\
19646  &  B   & 58193.5386&  35.944 \\ 
105080 &  A   & 58193.7546&  50.182 \\  
105080 &  A   & 58194.6308&  50.198 \\
105080 &  A   & 58195.6489&  50.179 \\
60845  &  BC  & 57985.4627&  42.462  \\
60845  &  BC  & 58193.7521&  42.524 \\ 
60845  &  BC  & 58194.6447&  42.514 \\
60845  &  BC  & 58195.6586&  42.523 \\
60845  &  BC  & 58177.7617&  42.513 \\
60845  &  BC  & 58232.5972&  42.540 \\
60845  &  BC  & 58242.5361&  42.513 \\
75663  &  A   & 57986.4885& $-$50.737 \\ 
75663  &  A   & 58177.8100& $-$56.446 \\
75663  &  A   & 58193.8257& $-$57.181 \\
75663  &  A   & 58195.8323& $-$57.305 \\
76816  &  B   & 57986.4996& $-$50.912 \\
76816  &  B   & 58193.8413& $-$50.916 \\ 
76816  &  B   & 58194.8267& $-$50.921  \\
76816  &  B   & 58195.8509& $-$50.912 \\  
78163  &  B   & 57986.5130&   6.116 \\  
78163  &  B   & 58194.8533&   7.152 \\ 
78163  &  B   & 58195.7872&   7.082 \\ 
78416  &  B   & 57986.5221&   0.752 \\
78416  &  B   & 58193.8595&   1.568 \\ 
78416  &  B   & 58194.8446&   1.816 \\ 
78416  &  B   & 58195.7774&   1.581 \\ 
80448  &  B   & 58193.8678&   7.509  \\
80448  &  B   & 58195.8008&   7.752  \\
80448  &  B   & 58194.8621&   7.679  \\
80448  &  B   & 58228.8113&   7.515 \\ 
80448  &  B   & 58233.8463&   7.353  \\
80448  &  B   & 58246.6693&   8.395  \\
80448  &  B   & 58248.8529&   8.176  \\
80448  &  B   & 58256.7395&   7.532  \\
80448  &  B   & 58257.8020&   7.916  \\
84789  &  B   & 57986.5346&   6.669 \\
84789  &  B   & 58195.8671&   5.260 
\enddata 
\end{deluxetable}

\begin{deluxetable*}{r l l rrr rr l}    
\tabletypesize{\scriptsize}     
\tablecaption{Position measurements and residuals (fragment)
\label{tab:speckle}          }
\tablewidth{0pt}                                   
\tablehead{                                                                     
\colhead{HIP/HD} & 
\colhead{System} & 
\colhead{Date} & 
\colhead{$\theta$} & 
\colhead{$\rho$} & 
\colhead{$\sigma_\rho$} & 
\colhead{(O$-$C)$_\theta$ } & 
\colhead{(O$-$C)$_\rho$ } &
\colhead{Ref.\tablenotemark{a}} \\
 & & 
\colhead{(yr)} &
\colhead{(\degr)} &
\colhead{(\arcsec)} &
\colhead{(\arcsec)} &
\colhead{(\degr)} &
\colhead{(\arcsec)} &
}
\startdata
 60845 & B,C &  1939.4600 &    357.8 &   0.3500 &   0.1500 &      4.2 &   0.1351 & M  \\
 60845 & B,C &  1956.3800 &    184.1 &   0.3100 &   0.0500 &      8.8 &   0.0956 & M \\
 60845 & B,C &  2018.1639 &     92.6 &   0.1769 &   0.0050 &      1.9 &   0.0070 & S \\
 60845 & A,BC &  1880.3800 &    270.5 &   2.4300 &   0.2500 &      2.2 &   0.3463 & M \\
 60845 & A,BC &  1991.2500 &    201.1 &   2.0510 &   0.0100 &     $-$1.2 &  $-$0.0042 & H \\
 60845 & A,B &  2018.1639 &    187.1 &   2.0965 &   0.0050 &      0.2 &  $-$0.0018 &  S
\enddata 
\tablenotetext{a}{
H: Hipparcos;
S: speckle interferometry at SOAR;
s: speckle interferometry at other telescopes;
M: visual micrometer measures;
G: Gaia DR2.
}
\end{deluxetable*}

\section{Individual objects}
\label{sec:obj}

For  each observed system,  the corresponding  Figure shows  a typical
double-lined CCF (the Julian date and components' designatios are marked
on the plot) together with the RV curve representing the orbit. In the
RV curves, squares denote  the primary component, triangles denote the
secondary component, while the full and dashed lines plot the orbit. 

\subsection{HIP 19639 and 19646 (triple)}

\begin{figure}
\epsscale{1.1}
\plotone{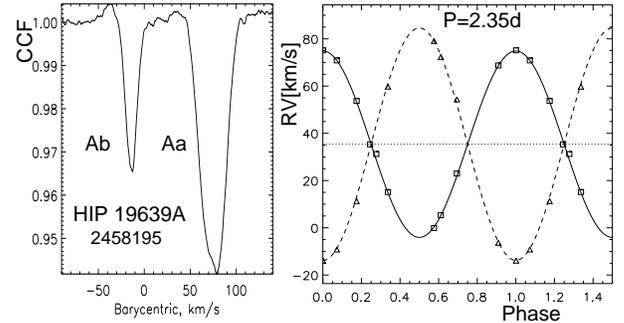}
\caption{CCF (left)  and RV curve (right) of HIP 19639 Aa,Ab.
\label{fig:19639}
}
\end{figure}

The 50\arcsec ~pair of bright stars HIP~19639 and 19646 was identified
as  a visual binary  by J.F.W.~Hershel  in 1838.   The {\it  Gaia} DR2
astrometry leaves no doubt that  this pair is physical: the components
have common  proper motion (PM),  distance, and RV. The  outer orbital
period  is  of the  order  of 240  kyr.   \citet{N04}  found that  the
component A is  a double-lined binary; its 2.35-day circular orbit is
determined  here  for the  first  time (Figure~\ref{fig:19639}).   Two
spectra (JD 2458052 and 2458053) were taken with the NRES spectrograph,
as described in \citep{paper3}.

The CCF of the stronger component Aa is wide and asymmetric, while the
CCF  of  Ab  is  narrower;  their  widths  correspond  to  approximate
projected rotation velocities $ V \sin i$ of 21.7 and 8.7 km~s$^{-1}$,
respectively, while the ratio of  the CCF areas implies $\Delta V_{\rm
  Aa,Ab} =  1.50$ mag. Wide and  shallow dips lead to  large RV errors
and large residuals to the orbit.  The mass ratio in the inner pair is
$q_{\rm Aa,Ab}  = 0.82$.  The RV of  the component B its  close to the
center-of-mass velocity  of A.   The component B  also has a  wide CCF
corresponding to $ V \sin i$ of $\sim$33 km~s$^{-1}$. 

\begin{figure}
\epsscale{1.1}
\plotone{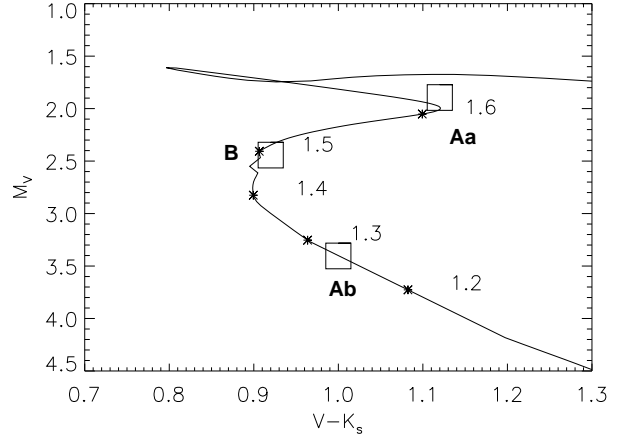}
\caption{Location of  three components of the HIP~19639  system on the
  color-magnitude  diagram  (squares).  The  full  line  is  a 2  Gyr
  isochrone for solar  metallicity \citep{Dotter2008}, where asterisks
  and numbers mark masses.
\label{fig:iso}
}
\end{figure}

Components  of   the  triple  system  HIP~19639  are   placed  on  the
color-magnitude  diagram  (CMD)  in  Figure~\ref{fig:iso},  using  the
distance modulus  of 5.47  mag. The  $V$ magnitudes of  Aa and  Ab are
estimated  from the  combined  magnitude of  A  and the  spectroscopic
magnitude difference of  1.5 mag. The $V-K$ color  of Ab, not measured
directly, is  assumed to place it  on the main sequence.   It is clear
that  both Aa  and B  are located  above the  main sequence,  near the
turn-off. Their  positions match reasonably  well the 2  Gyr isochrone
and correspond to the masses of 1.6 and 1.5 ${\cal M}_\odot$. The mass
of Ab  deduced from the  isochrone is 1.28 ${\cal  M}_\odot$, matching
the spectroscopic mass ratio, while the radii of Aa and Ab are 2.7 and
1.1 $R_\odot$.  The orbital axis $a_1 + a_2 = 10.6 R_\odot$ means that
the  binary  is detached.   However,  contact  and  mass transfer  are
imminent when Aa expands further.

The spectroscopic  mass sum  of the inner  binary is only  0.18 ${\cal
  M}_\odot$.   The mass  sum estimated  above, 2.88  ${\cal M}_\odot$,
implies  an inclination  $i_{\rm Aa,Ab}  = 23\fdg4$,  or  $\sin i_{\rm
  Aa,Ab} = 0.40$, hence the  synchronous rotation velocities of Aa and
Ab are 23.8  and 9.7 km~s$^{-1}$, respectively, in  agreement with the
measured CCF width.  Summarizing, this is an interesting triple system
where the inner close binary is caught at evolutionary phase preceding
the mass transfer.

\subsection{HD 105080 and 105081 (2+2 quadruple) }

\begin{figure}
\epsscale{1.1}
\plotone{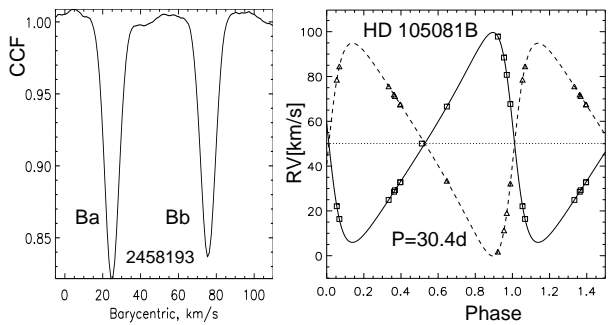}
\caption{CCF (left)  and RV curve (right) of HD 105081 Ba,Bb.
\label{fig:105080}
}
\end{figure}

Two nearly equal stars HD~105080  and 105081 form a 12\farcs9 physical
binary, first measured by J.~Hershel  in 1835.  Each of these stars is
a close pair, making the whole  system quadruple.  The pair Aa,Ab is a
known visual binary RST~4958  with a small magnitude difference. Since
its discovery in 1942 by R. A. Rossiter, it was observed only episodically.
By adding  three speckle  measures made  at SOAR in  2017 and  2018, a
preliminary  orbit  with  $P  =  91.6$  years  can  be  fitted  to  the
observations (Table~\ref{tab:vborb}).  Double  lines were reported for
this star in the literature,  although there could be a confusion with
the double-lined component  B.  The RV of A  is practically coincident
with the center-of-mass velocity of B.  The {\it Gaia} DR2 parallax of
A has  a large error of 1\,mas,  being affected by the  Aa,Ab pair.  I
adopt  the  parallax  of  B  as  the distance  to  the  system.   Both
components are then located on the CMD very close to each other, above
the  main  sequence. This  distance  and  the  visual orbit  of  Aa,Ab
correspond to the mass sum of 1.7 ${\cal M}_\odot$; however, the orbit
is poorly constrained.

The component B (HD~105081), which  is only slightly fainter than A in
the $V$ and $G$ bands, is revealed here to be a twin double-lined pair
with      $P=30.4$       days      and      eccentricity      $e=0.42$
(Figure~\ref{fig:105080}).  The spectroscopic mass sum of
Ba and Bb, 1.98 ${\cal M}_\odot$, and the mass sum inferred from the
absolute magnitudes, 2.26 ${\cal M}_\odot$, imply inclination
$i_{\rm Ba,Bb}=  73\degr$. The CCF  dips of Ba  and Bb are  narrow and
deep,   hence   the  residuals   to   the   orbit   are  small,   only
0.07~km~s$^{-1}$ .

\subsection{HIP 60845 (quintuple)}

\begin{figure}
\plotone{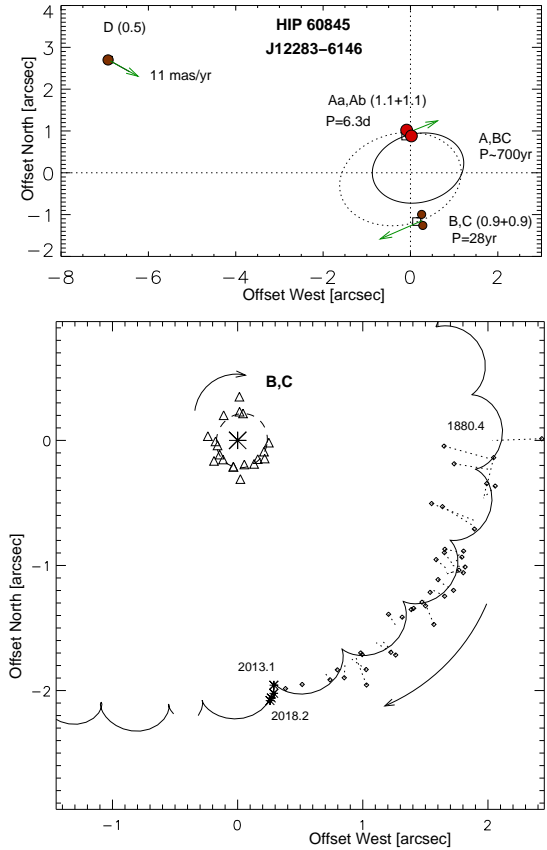}
\caption{Quintuple system HIP 60845 (WDS J12283$-$6146). The positions
  of three  components A, BC, and D  on the sky and  their motions are
  illustrated in  the upper panel.  Periods and  masses are indicated.
  The lower panel shows the observed motion of the subsystems A,BC and
  B,C and their  orbits. In this plot, the  coordinate origin coincides
  with  the main  star  A, the  wavy  line shows  the  motion of  the
  component B  around A  according to the  orbit. Small  crosses depict
  measurements  of A,BC where  the pair  BC was  unresolved, asterisks
  depict the resolved  measurents of A,B. The orbit  of B,C is plotted
  on the  same scale around the   coordinate origin  by the dashed
  line and triangles. 
\label{fig:HIP60845}
}
\end{figure}

\begin{figure}
\epsscale{1.1}
\plotone{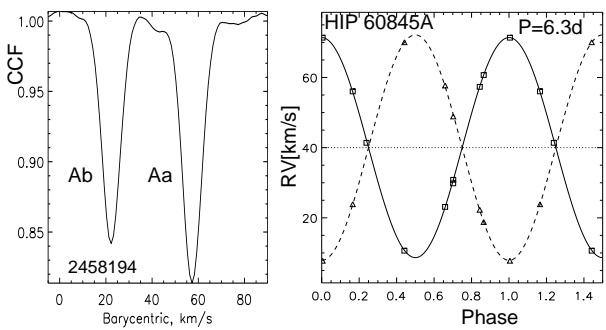}
\caption{CCF (left)  and RV curve (right) of HIP 60845 Aa,Ab. 
\label{fig:60845}
}
\end{figure}

This  multiple  system is  located  within  50\,pc  from the  Sun  and
contains  at least  five stars  arranged  in a  rare 3-tier  hierarchy
illustrated  in Figure~\ref{fig:HIP60845}.   The widest  7\farcs9 pair
RST~4499 AB,D is  physical, based on its stable  relative position and
the {\it  Gaia} DR2  parallaxes and PMs  of the components.   The fast
(175 mas~yr$^{-1}$) PM facilitates discrimination between physical and
optical companions, despite the high stellar density in this field and
the faintness  of D ($V  = 13.4$ mag).   The period of  AB,D estimated
from its projected separation $\rho_{\rm AB,D} = 7\farcs88$ is about 4
kyr and corresponds to the  chracteristic orbital velocity  $\mu^* =
2 \pi \rho_{\rm AB,D}/P_{\rm  AB,D} = 13$ mas~yr$^{-1}$.  The relative
PM  between A  and D,  measured by  {\it Gaia}  and corrected  for the
orbital  motion of  A,  is  11 mas~yr$^{-1}$;  it  is directed  almost
exactly toward A (position angle  240\degr), as indicated by the arrow
in  Figure~\ref{fig:HIP60845}. If D  moves on  an eccentric  orbit, it
will come  close to  A,BC in $\sim$700  years, disrupting  the system.
Alternatively,  the  observed  motion  might correspond  to  a  highly
inclined and not very eccentric  outer orbit, in which case the system
could  be dynamically stable.   If the  pair AB,D  is bound,  the true
separation between A  and D cannot exceed its  projected separation by
more   than  $\sim$2  times,   given  their   relative  speed   of  11
mas~yr$^{-1}$ and the total mass sum of 4.5 ${\cal M}_\odot$.

The  visual  binary  A,BC  (CPO~12),  for which  a  crude  orbit  with
$P=2520$\,years  and semimajor  axis of  5\farcs4 has  been  published by
\citet{USN2002}, occupies  the intermediate hierarchical  level.  This
orbit  is poorly  constrained  by the  century-long  observed arc.   I
computed an  alternative orbit  with $P_{\rm A,BC}  = 690$ years,  with 
smaller eccentricity and semimajor axis (Table~\ref{tab:vborb}).  This
orbit makes more sense, given  the threat to dynamical stability posed
by the  outer component D.  Even  then, the period  ratio $P_{\rm AB,D}/
P_{\rm  A,BC}  \sim  5$  is  comparable  to  the  dynamical  stability
limit. On the other hand, the  nearly circular orbit of the inner pair
B,C (RST~4499) with $P=28.2$\,years  is definitive.  The visual orbits and
the estimated mass  sums match the {\it Gaia}  DR2 parallax reasonably
well.  Both orbits are retrograde and have  small inclinations.

The visual primary star A  was identified as a spectroscopic binary by
\citet{N04}.   Now  its  6.3-day  double-lined  orbit  is  determined
(Figure~\ref{fig:60845}).  The eccentricity  does not differ from zero
significantly, hence the circular orbit  is imposed.  The masses of Aa
and  Ab are  almost equal,  as are  their CCF  dips.  Given  the small
angular distance  between A and BC,  2\farcs06, the light  is mixed in
the  fiber,  so   the  CCF  often  has  3  dips;   the  CCF  shown  in
Figure~\ref{fig:60845} is  an exception recorded with  good seeing and
careful guiding.  The  magnitude difference between Ab and  Aa is 0.37
mag, hence their individual $V$  magnitudes are 7.79 and 8.16 mag.  By
comparing  the mass sum  of Aa  and Ab  estimated from  their absolute
magnitudes, 2.2 ${\cal M}_\odot$,  with the spectroscopic mass sum of
0.167 ${\cal  M}_\odot$, I find that  the orbit of Aa,Ab  has a small
inclination  of  $i_{\rm  Aa,Ab}  \approx 25\degr$.   The  synchronous
rotation  of  the component  Aa,  of  one  solar radius,  implies  the
projected  rotation of  $V \sin  i =  3.3$ km~s$^{-1}$,  close  to the
measured value.  The three inner orbits could be close to coplanarity,
given their small inclinations. 

The CCF  dip corresponding to the  combined light of BC  is narrow and
has a  constant RV  of 42.5 km~s$^{-1}$,  close to  the center-of-mass
velocity of  A.  Slow  axial rotation, location  of components  on the
main  sequence in  the  CMD,  and non-detection  of  the lithium  line
suggest that this quintuple system is relatively old.

\subsection{HIP 61840 (binary)}

\begin{figure}
\epsscale{1.1}
\plotone{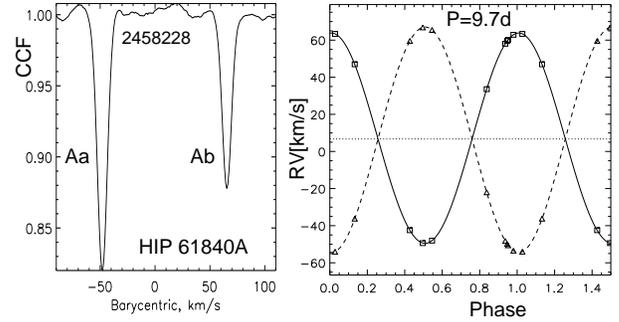}
\caption{CCF (left)  and RV curve (right) of HIP 61840 Aa,Ab.
\label{fig:61840}
}
\end{figure}

Unlike  the rest  of the objects  in  this paper,  this star  is a  simple
spectroscopic binary without additional  components. It belongs to the
67 pc  sample  of solar-type  stars  \citep{FG67a}  and  is young,  as
inferred from the chromoshperic activity, X-ray flux, and the presence
of lithium  in its atmosphere.  The double-lined  nature was announced
by \citet{Wichman2003} and \citet{N04}, but the orbital period was, so
far, unknown.  The object  has been observed by speckle interferometry
at SOAR in 2011 and 2016 and found unresolved.

The orbit with  $P=9.67$ days has a small,  but significantly non-zero
eccentricity   $e=  0.007 \pm  0.001$  (Figure~\ref{fig:61840}). The
residuals to the circular orbit are 0.3 km~s$^{-1}$,  6$\times$
larger than to the eccentric orbit.  The masses of Aa and Ab estimated
from  absolute magnitudes  are  1.24 and  1.13  ${\cal M}_\odot$,  the
spectroscopic mass sum is 1.62 ${\cal M}_\odot$, hence the inclination
is $i =  62\degr$.  The measured projected rotation  speed matches the
synchronous speed.

\subsection{HIP 75663 (2+2 quadruple)}

\begin{figure}
\epsscale{1.1}
\plotone{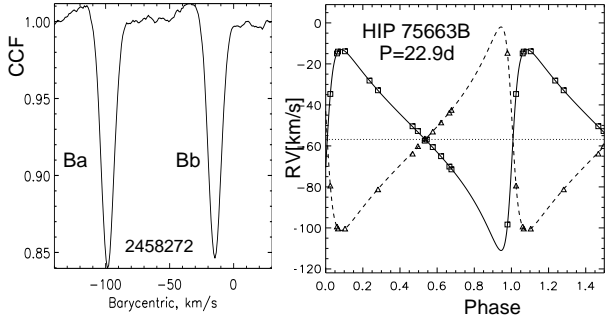}
\caption{CCF (left)  and RV curve (right) of HIP 75663 Ba,Bb.
\label{fig:75663}
}
\end{figure}

This system is  quadruple.  The outer 9\farcs4 pair  was discovered in
1825 by W.~Struve.  Its estimated period is 17\,kyr. \citet{N04} noted
double lines  in the visual secondary  B.  Its orbital  period is 22.9
days (Figure~\ref{fig:75663}),  with a large  eccentricity of $e=0.61$
and the mass ratio $q_{\rm Ba,Bb}  = 0.997$ (a twin). The areas of the
CCF dips  of Ba and Bb are  equal to within 2\%.  The masses estimated
from absolute  magnitudes are 1.02  ${\cal M}_\odot$ each,  leading to
the  orbital  inclination   of  $i_{\rm  Ba,Bb}=55\degr$.   The  axial
rotation of Ba and Bb is faster than synchronous, as expected for such
eccentric orbit.

The RV  of the main  component A is  variable according to  the CHIRON
data (range  from $-$50.7 to $-$57.3 km~s$^{-1}$)  and the literature.
\citet{N04} made two measurements averaging at $-58.2$ km~s$^{-1}$ and
suspected   RV  variability,   while  {\it   Gaia}   measured  $-55.4
\pm 2.$ km~s$^{-1}$.   The photo-center motion  of A  caused by  the subsystem
Aa,Ab  could  explain  the  discrepancy  between the  {\it  Gaia}  DR2
parallaxes of A and B {\bf (9.29$\pm$0.16 and 7.69$\pm$0.07 mas,
  respectively). A similar discrepancy of parallaxes exists in the HD
  105080/81 system, where A is a visual binary. } The period of Aa,Ab is not known; presumably it
is longer  than a year.  \citet{Isaacson2010} classified  this star as
chromospherically    active    and    found    the   RV    jiter    of
4.2\,m~s$^{-1}$. The location of the  component A on the CMD indicates
that  it  is slightly  evolved  and  matches  approximately the  4-Gyr
isochrone. Lithium is detectable in the spectra of A and B.

\subsection{HIP 76816 (2+2 quadruple)}

\begin{figure}
\epsscale{1.1}
\plotone{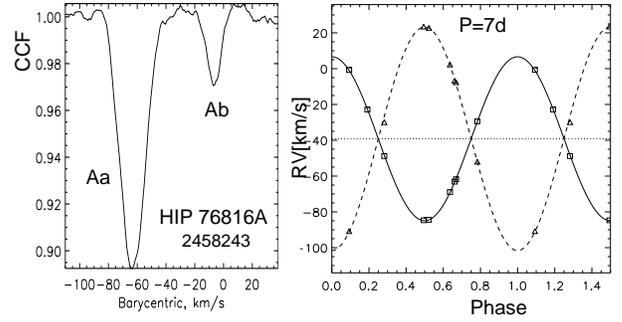}
\caption{CCF (left)  and RV curve (right) of HIP 76816 Aa,Ab.
\label{fig:76816}
}
\end{figure}

This  is a  quadruple system  located at  309\,pc from  the  Sun.  The
5\farcs4  visual binary  HWE~37  is known  since  1876; its  estimated
period is 33\,kyr.  Double lines in the component A were discovered by
\citet{Desidera2006}. I used their measurement to refine the period of
the  circular  orbit  of  Aa,Ab  with $P=6.95$  days,  determined  here
(Figure~\ref{fig:76816}).  The eccentric  orbit has similar residuals,
hence the circular solution is retained.  The components Aa and Ab are
unequal in the amplitudes of their  CCF dips (area ratio 0.152, or 2.0
mag difference) and of the RV  variation (mass ratio 0.735). The RV of
the visual  component B  is also variable  with a long,  still unknown
period.  I  measured its RV  at $-$50.9 km~s$^{-1}$  (constant), while
{\it  Gaia}  measured  $-41.4$  km~s$^{-1}$  and  \citet{Desidera2006}
measured $-37.5$ km~s$^{-1}$; these RVs differ from the center-of-mass
velocity of A, $-39.94$ km~s$^{-1}$.

The matching  {\it Gaia} DR2 parallaxes  place both A and  B above the
main sequence. The component B is  more evolved: it is brighter than A
in the  $K$ band (unless  its $K$ magnitude  measured by 2MASS  is 
distorted  by  the  proximity  of  A,  as  happens  with  other  close
pairs). The mass sum of Aa and Ab, estimated crudely from the absolute
magnitudes,  is almost 3  ${\cal M}_\odot$,  leading to  the
orbital inclination  of 42\fdg5.    The stars
Aa and Ab apparently rotate synchronously with the orbit.

\subsection{HIP 78163 (2+2 quadruple)}

\begin{figure}
\epsscale{1.1}
\plotone{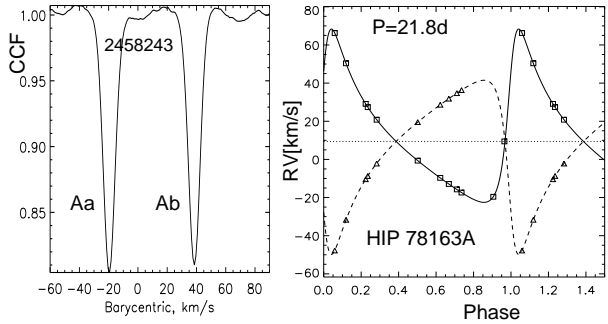}
\caption{CCF (left)  and RV curve (right) of HIP 78163 Aa,Ab.
\label{fig:78163}
}
\end{figure}

This multiple  system is  composed by the  outer 7\farcs9  pair WG~185
(estimated period 8\,kyr) and  the inner subsystem Aa,Ab discovered by
\citet{N04}. For  the latter,  I determined here  the orbit  with $P=
21.8$ days (Figure~\ref{fig:78163}),  $e=0.58$, and mass ratio $q_{\rm
  Aa,Ab} = 0.996$  (a twin).  The RV of the  component B measured with
CHIRON  ranges  from 6.1  to  7.1  km~s$^{-1}$  and differs  from  the
center-of-mass RV of the  component A, 9.4 km~s$^{-1}$.  Considering
also  the {\it  Gaia} RV(B)=16.3  km~s$^{-1}$, I  infer that  B  is a
single-lined  binary, possibly  with  a  long period  and  a small  RV
amplitude.   Its  photo-center motion  could  explain  the slight  
 discrepancy between  {\it Gaia} parallaxes and PMs of A
and B. Therefore, the parallax of A, 10.42\,mas, is likely the correct
one.  

The  twin  components  Aa  and  Ab  have masses  of  one  solar  each.
Comparting  them  to $M  \sin^3  i$,  the  inclination of  50\degr  ~is
derived.   The   stars   A   and   B   are   located   on   the   main
sequence. Interestingly,  lithium is detectable  in the spectra  of Aa
and Ab, but not in B, which is a similar solar-mass star.

\subsection{HIP 78416 (triple or quadruple)}

\begin{figure}
\epsscale{1.1}
\plotone{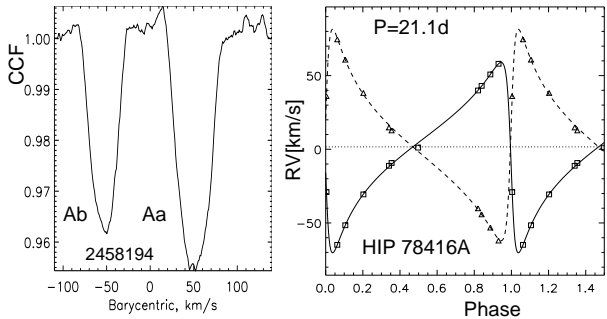}
\caption{CCF (left)  and RV curve (right) of HIP 78416 Aa,Ab.
\label{fig:78416}
}
\end{figure}

The outer  6\farcs5 pair  HWE~81, known since  1876, has  an estimated
period  of  10\,kyr.   \citet{N04}  detected  RV  variability  of  the
component   A,  later   found   to  be   a   double-lined  binary   by
\citet{Desidera2006}.   The   orbital  period   is  21  days   and  the
eccentricity  $e=0.766$ is  unusually  high for  such  a short  period
(Figure~\ref{fig:78416}).  The  wide and  shallow CCF dips  imply fast
axial  rotation.  For  this  reason,  the RVs  are  not measured  very
accurately  and the  residuals to  the orbit  are large,  0.2  and 0.6
km~s$^{-1}$.  By comparing the estimated masses of Aa and Ab, 1.15 and
1.03 ${\cal M}_\odot$ respectively, with  $M \sin^3 i$, I estimate the
orbital inclination of 74\degr.

The visual component B  also has a fast axial rotation of  $V \sin i \sim
40$ km~s$^{-1}$, degrading the accuracy of its RV measurement.  The
RVs of B measured with CHIRON, by {\it Gaia}, and by \citet{Desidera2006}
(1.4, $-$0.7,  and 0.5 km~s$^{-1}$ respectively)  are reasonably close
to  the center-of-mass  RV of  A, 1.7  km~s$^{-1}$. Therefore,  B is
likely a  single star. All three  stars Aa, Ab, and  B have comparable
masses and  similar colors.  The component A,  being a close  pair, is
located on the CMD just above B, as expected.

The RVs of Aa and Ab measured by \citet{Desidera2006}, 58.9 and $-1.8$
km~s$^{-1}$,  correspond  to   the  center-of-mass  velocity  of  30.2
km~s$^{-1}$  and do  not fit  the present  orbit with  $\gamma  = 1.7$
km~s$^{-1}$.   This  discrepancy suggests  that  Aa,Ab  is orbited  by
another close  companion.  Further  monitoring is needed,  however, to
prove this hypothesis. 

According to  \citet{Rizzuto2011}, this system belongs to  the Sco OB2
association with  a probability of 74\%. Fast  axial rotation and the
presence of lithium indicate a young age.

\subsection{HIP 80448 (2+2 quadruple)}

This young multiple system is  located within 50\,pc from the Sun.  It
contains  four components  in  a  small volume.   The  outer pair  A,B
(COO~197) has  an uncertain visual orbit with  a millenium-long period
\citep{Ary2015b}.  Its secondary component was resolved in 2004 into a
0\farcs13  pair CVN~27  Ba,Bb by  \citet{Chauvin2010},  using adaptive
optics.   Independently,   a  subsystem  TOK~50   Aa,Ab  with  similar
separation  was  discovered in  2009  by  \citet{TMH10} using  speckle
interferometry.  In  fact,  the   same  subsystem  Ba,Bb  was  wrongly
attributed to  the primary component;  its published measures  at SOAR
with  angle inverted  by 180\degr  ~match the  preliminary  orbit with
$P_{\rm Ba,Bb}=20$ years.  The pair  TOK~50 Aa,Ab does not exist.  The
component Bb is fainter than Ba by  3.5 mag in the $V$ band and by 1.1
mag  in the  $K$ band;  its  lines are  not detected  in the  combined
spectrum of all stars.

\begin{figure}
\plotone{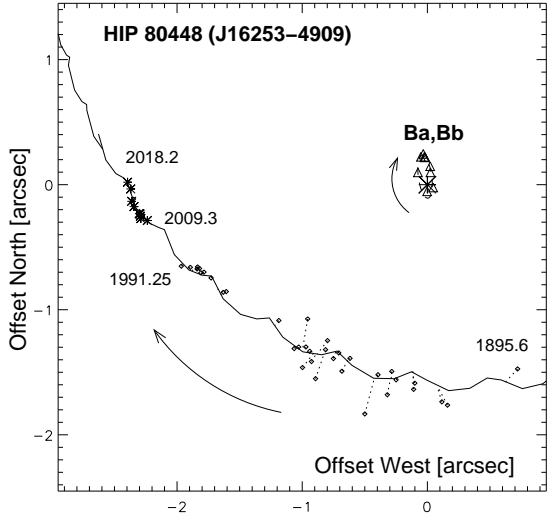}
\caption{Visual orbits of HIP~80448  A,B and Ba,Bb (WDS J16253$-$4909,
  COO~197 and CVN~27).
\label{fig:COO197}
}
\end{figure}

\begin{figure}
\epsscale{1.1}
\plotone{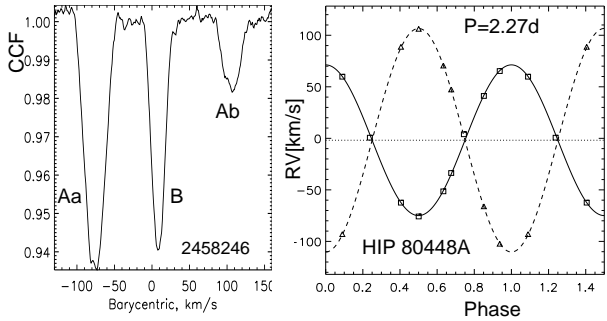}
\caption{CCF (left)  and RV curve (right) of HIP 80448 Aa,Ab. 
\label{fig:80448}
}
\end{figure}

Figure~\ref{fig:COO197} shows the  positions of resolved components on
the sky and the fitted orbits. Considering that the outer orbit is not
constrained by  the data, I fixed  its period to $P_{\rm  A,B} = 1950$
years  and  its  axis to  $a_{\rm  A,B}  =  4\farcs64$ to  adjust  the
dynamical mass sum to its  estimated value, 3.5 ${\cal M}_\odot$.  The
orbit of Ba,Bb with $P_{\rm Ba,Bb}  = 20$ years yields the mass sum of
1.8 ${\cal  M}_\odot$, close to  the estimated one. The  ``wobble'' in
the positions  of A,Ba  caused by the  subsystem is clearly  seen. Its
relative amplitude  $f = 0.40$  corresponds to the mass  ratio $q_{\rm
  Ba,Bb} = f/(1-f) = 0.67$ that argees with the magnitude difference.

The spectrum  of the main component  A (in fact, blended  light of Aa,
Ab, and Ba) shows stationary lines of Ba and double lines of Aa and Ab
in  rapid  motion; the  subsystem  Aa,Ab  was  discovered with  CHIRON
\citep{survey}.  It is found here  that the orbital period  is
$P_{\rm   Aa,Ab}   =   2.3$    days   and   the   orbit   is   circular
(Figure~\ref{fig:80448}).  The  mass ratio  $q_{\rm Aa,Ab} =  0.67$ is
similar  to the mass  ratio $q_{\rm  Ba,Bb}$, while  the ratio  of dip
areas corresponds to $\Delta V_{\rm  Aa,Ab} = 1.9$ mag.  Comparison of
estimated and spectroscopic mass sums leads to the orbital inclination
$i_{\rm Aa,Ab}  = 66\degr$  or $i_{\rm Aa,Ab}  = 114\degr$. It  is not
dissimilar to the inclination of   other inner pair, $i_{\rm Ba,Bb}
= 109\degr$, but it is  difficult to believe that these two subsystems
have coplanar orbits, given the huge difference of their periods.

The  components Aa  and Ab  rotate synchronously  with the  orbit. The
projected rotation of Ba, $V \sin i = 14.3$ km~s$^{-1}$, is relatively
fast, supporting the thesis that this system is young. The presence of
lithium also suggests  youth.  The four components are  located in the
CMD at about 0.7 mag above the main sequence.

The  pair  Ba,Bb is  presently  near  the  periastron of  its  20 year
orbit. I  measured the  RV(Ba) from 7.51  to 8.18  km~s$^{-1}$, quite
different from the center-of-mass  velocity of A, $-2.08$ km~s$^{-1}$.
This positive  difference and the positive trend actually  match the orbit
of  Ba,Bb; I predict  that RV(Ba)  will soon  start to  decrease. The
orbits  of  A,B  and  Ba,Bb   are  not  coplanar,  although  both  are
retrograde.

\subsection{HIP 84789 (triple)}

\begin{figure}
\epsscale{1.1}
\plotone{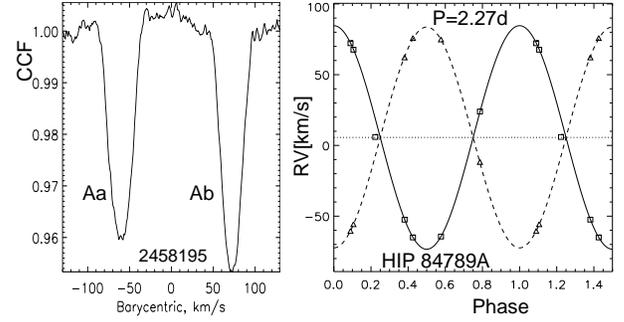}
\caption{CCF (left) and RV curve (right) of HIP 84789 Aa,Ab. Note that
  the secondary component Ab has a dip with larger amplitude.
\label{fig:84789}
}
\end{figure}

This 5\farcs6 visual binary STF~2148, discovered in 1832 by W.~Struve,
has  an estimated  period of  17\,kyr.   Double lines  in its  primary
component  A were  noted by  \citet{N04}.  The  orbital period  of the
subsystem Aa,Ab determined here is $P_{\rm  Aa,Ab} = 2.3$ days; it is a
twin pair  with $q_{\rm  Aa,Ab} =  0.988$ (Figure~\ref{fig:84789}). The  deeper CCF
dip belongs to the less massive component Ab; the more massive star Aa
rotates a  little faster and has the  dip area 3\% larger  than Ab, as
well  as the  smaller RV  amplitude. The  RVs of  both  components are
measured  with large  errors owing  to the  wide and  low-contrast CCF
dips; the residuals  to the orbits are also large.   An attempt to fit
the  orbit  with non-zero  eccentricity  does  not  result in  smaller
residuals, hence the orbit is circular.

The  estimated masses of  Aa and  Ab are  1.30 ${\cal  M}_\odot$ each,
leading   to   the   orbital   inclination   of   $i_{\rm   Aa,Ab}   =
45\degr$.  Assuming  the  radii  of  1.3  $R_\odot$,  the  synchronous
rotation velocity  is  $V  \sin  i  =  20.2$  km~s$^{-1}$.  The  width  of  the
correlation dips  matches this estimate and suggests  that Aa rotates
slightly faster and Ab slightly slower than synchronous.
 
The two  components A and  B are located  on the CMD above  each other
(they have  the same  color) because A  contains two equal  stars; the
mass of B is  very similar to the masses of Aa  and Ab, 1.3 solar. The
component B is single, as inferred  from the equality of its RV to the
center-of-mass velocity of  A. However, it rotates much  faster, at $V
\sin i  \sim 49$ km~s$^{-1}$. Very likely,  the rotation of Aa  and Ab has
been slowed down by tidal synchronization with the orbit.

\section{Summary}
\label{sec:sum}

\begin{figure}
\plotone{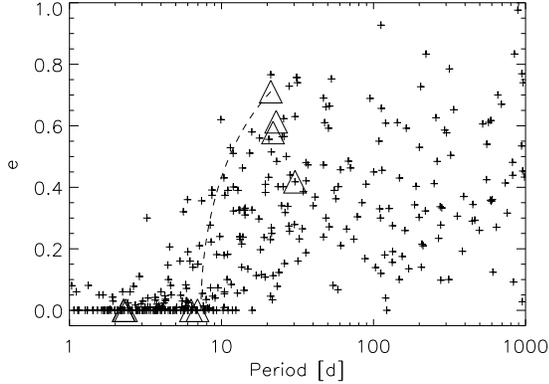}
\caption{Eccentricity vs.  period for members  of hierarchical systems
  studied here  (large triangles)  and for 467  spectroscopic binaries
  from  the   MSC  with   primary  masses  from   0.5  to   1.5  solar
  (crosses). The  dashed line shows  the locus of HIP~78416  Aa,Ab for
  evolution with  constant angular momentum,  $P(1 - e^2)^{3/2}  = {\rm
    const}$.
\label{fig:pe}
}
\end{figure}

Probably by  accident, the periods  of 9 spectroscopic  systems within
hierarchical  multiples  are equally  divided  between three  distinct
groups:  (i)   circular  orbits  with   $P  \approx  2.3$   days,  (ii)
intermediate  periods  between  6  and  9  days,  circular  or  nearly
circular, and (iii) eccentric orbits  with periods from 21 to 30 days.
Figure~\ref{fig:pe}  places these  orbits  on the  period-eccentricity
plot.   The plus  signs are  467 spectroscopic  binaries  with primary
masses from 0.5 to 1.5 ${\cal M}_\odot$ from the MSC \citep{MSC}. When
the  eccentric orbits  of the  group (iii)  are  tidally circularized,
their periods  will match those  of group (ii), suggesting  that these
subsystems could be formed by  a common mechanism, such as Kozai-Lidov
cycles with dynamical tides \citep{Moe2018}.  The periods of group (i)
are  substantially  shorter,  so  their  formation  history  could  be
different.

Six out  of the 10 double-lined  binaries studied here  are twins with
mass  ratio  $q >  0.95$,  while the  lowest  measured  mass ratio  is
0.67. If  the mass ratios  were uniformly distributed in  the interval
(0.7, 1.0), where double lines  are detectable, the fraction of twins
would be only 0.15, whereas in  fact it is 0.6.  It is established that
twins correspond to a well-defined peak in the mass ratio distribution
of solar-type spectroscopic binaries \citep{twins}.  They are believed
to be formed when a low-mass binary accretes a major part of its mass.
The mass  influx also creates  conditions for formation  of additional
components, building  stellar hierarchies ``from  inside out''.  Thus,
twins are  naturally produced as inner components  of multiple systems
in the process of mass assembly.

The  goal  of   this  study  was  to  determine   unknown  periods  of
spectroscopic subsystems in several multiple stars. Although this goal
is reached,  I discovered RV  variability of other  visual components
(HIP 75663A,  76816B, and 78163B),  converting these triples  into 2+2
quadruples.  The periods  of new  subsystems, presumably  long, remain
unknown so far.

\acknowledgements

I thank the operator of  the 1.5-m telescope R.~Hinohosa for executing
observations  of  this  program  and  L.~Paredes  for  scheduling  and
pipeline processing.  Re-opening of CHIRON  in 2017 was largely due to
the enthusiasm and energy of T.~Henry.

This work  used the  SIMBAD service operated  by Centre  des Donn\'ees
Stellaires  (Strasbourg, France),  bibliographic  references from  the
Astrophysics Data  System maintained  by SAO/NASA, and  the Washington
Double Star Catalog maintained at USNO.

This work has made use of data from the European Space Agency (ESA) mission
{\it Gaia} (\url{https://www.cosmos.esa.int/gaia}), processed by the {\it Gaia}
Data Processing and Analysis Consortium (DPAC,
\url{https://www.cosmos.esa.int/web/gaia/dpac/consortium}). Funding for the DPAC
has been provided by national institutions, in particular the institutions
participating in the {\it Gaia} Multilateral Agreement.

\facilities{CTIO:1.5m, SOAR}






\end{document}